# Optimizing Variable Flip-Angles in Magnetization-Prepared Gradient Echo Sequences for Efficient 3D-T1ρ Mapping

Marcelo V W Zibetti[1], Hector L. De Moura[1], Mahesh B. Keerthivasan[2], and Ravinder R Regatte[1]

[1] Center for Biomedical Imaging, Department of Radiology, New York University Grossman School of Medicine, New York, NY, USA
[2] Siemens Medical Solutions USA Inc, Malvern, PA.

## Abstract

**Purpose**: To optimize the choice of the flip-angles of magnetization-prepared gradient echo (MP-GRE) sequences for improved accuracy, precision, and speed of 3D-$T_{1\rho}$ mapping.

**Methods**: We propose a new optimization approach for finding variable flip-angle (VFA) values that improve MP-GRE sequences used for 3D-$T_{1\rho}$ mapping. This new approach can simultaneously improve the accuracy and signal-to-noise ratio (SNR) while reducing filtering effects. We demonstrate the concept in the three different versions of the MP-GRE sequences that are typically used for 3D-$T_{1\rho}$ mapping and evaluate their performance in model agarose phantoms (n=6) and healthy volunteers (n=5) for knee joint imaging. We also tested the optimization with sequence parameters targeting faster acquisitions.

**Results:** Our results show that optimized VFA can improve the accuracy and the precision of the sequences, seen as a reduction of the mean of normalized absolute difference (MNAD) from 6~8% to 4~5% in model phantoms and from 14~22% to 12~14% in the knee joint, and improving SNR from 12~27 to 24~35 in agarose phantoms and 5~13 to 11~16 in healthy volunteers. The optimization can also compensate for the loss in quality caused by making the sequence faster. This results in sequence configurations that acquire nearly twice more data per unit of time with similar SNR and MNAD measurements as compared to its slower versions.

**Conclusion:** The optimization of the VFA can be used to increase accuracy and precision, and improve the speed of the typical imaging sequences used for quantitative 3D-$T_{1\rho}$ mapping of the knee joint.

# 1. INTRODUCTION

The spin-lattice relaxation time in the rotating frame ($T_{1\rho}$) has shown to be useful in several applications, ranging from quantifying proteoglycan content of the cartilage (1,2) to distinguishing between healthy and chronic liver disease subjects (3) and with Alzheimer's and Parkinson's disease (4,5). One drawback is the undesirably long acquisition time required for $T_{1\rho}$ mapping, particularly when three-dimensional (3D) and high-resolution (HR) mapping is desired. High precision and reduced acquisition time are fundamental for clinical use, improving diagnostic quality while avoiding long exams.

Significant efforts have been made to reduce scan time while maintaining good image quality. Fast pulse sequences such as fast gradient-echo (6–8) and fast spin-echo (9), and sequences with long readouts, such as echo-planar imaging (10) and spiral imaging (11), are examples of efficient sequences that capture more k-space data per unit of time. Another approach to reducing the scan time is undersampling. In parallel imaging (PI) (12–14), multiple receiving coils are used to capture more data in parallel, with different coil sensitivities, allowing for undersampling of the k-space to reduce acquisition time. Compressed sensing (CS) (15–17), which relies on incoherent sampling and sparse reconstruction, also reduces time by undersampling, while obtaining high-quality images due to the advanced non-linear reconstruction. Recent advancements with data-driven approaches for learned reconstruction and sampling pattern (SP) have shown that k-space sampling can be optimized for specific anatomy and reconstruction methods (18–22) for improved quality in accelerated MRI. All these improvements can be combined, allowing 3D MRI with short scan times. This study focuses on fast pulse sequences for quantitative mapping.

The 3D-$T_{1\rho}$ mapping methods discussed in this study are based on Cartesian magnetization-prepared gradient-echo (MP-GRE) pulse sequences (23) with RF and gradient spoilers (24), modified to include $T_{1\rho}$ preparation (25–28). In these sequences, a train of small flip-angle RF pulses combined with the action of the magnetic gradient system creates multiple gradient echoes that correspond to the acquisition of multiple k-space lines after a $T_{1\rho}$ preparation module. Although very efficient, in the sense that multiple k-space lines are acquired in each echo train, it has still limitations due to the loss of the prepared magnetization that happens due to the natural signal evolution (SE) during the transient state of the echo train (25,26,28). This phenomenon cause issues such as spin-lattice relaxation time ($T_1$) contamination and k-space filtering effect (FE) (25,29), ultimately affecting the accuracy of measured $T_{1\rho}$ values.

To minimize k-space FE, several strategies have been implemented, such as the use of variable flip-angle (VFA) (26,28), the use of a limited number of echoes per train, also called view per segments (VPS), and RF phase cycling (28). These techniques are combined with k-space data collection ordering schemes that capture first phase/partition encoding positions near the center of k-space (28,30,31), preserving most of the desired contrast. RF phase cycling incurs a time penalty by doubling the scan time because two acquisitions with opposite longitudinal magnetization are used to reduce $T_1$ contamination. Limiting the VPS captured also reduces significantly k-space FEs, but usually leads to an increase in total scan time. To balance between k-space FE and scan time due to multiple shots, most sequences use from 64 to 256 VPS. On the other hand, the use of VFA and proper k-space ordering incurs no time penalty and can be exploited to reduce FE (26,28).

Besides FEs and $T_1$ contamination, noise is also an important factor that affects the quality of $T_{1\rho}$ mapping. One way to reduce the sensitivity to noise is to improve the choice of spin-lock times (TSLs) (32–34). Using optimized choices for TSLs improves the quality of $T_{1\rho}$ mapping, allowing a reduction of the number of TSLs with the same quality. Another way to improve $T_{1\rho}$ mapping is to modify the pulse sequence to increase signal strength, obtaining a better SNR. Improved SNR can be obtained by using larger flip-angles (FAs) combined with longer recovery times. While larger FAs may be beneficial, longer recovery times usually lead to an increase in total scan time.

In this work, we propose an innovative data-driven optimization framework for $T_{1\rho}$ mapping with MP-GRE sequences. Inspired by (26,28), we optimized the VFA to improve the sequence. However, different from previous works, we used a target cost function that involves three components: 1) $T_{1\rho}$ mapping accuracy, 2) flatness of the SE (to minimize FE), and 3) signal strength (for improved SNR). The proposed optimization framework is applied to three different pulse sequences typically used for $T_{1\rho}$ mapping: 1) The magnetization-prepared angle-modulated partitioned k-space spoiled GRE snapshots (MAPSS) sequence (28); 2) the tailored VFA MP-GRE sequences with magnetization reset (26), called here MP-GRE-WR, and 3) standard MP-GRE sequences (27,35). As it will be shown later, the proposed framework can generalize the choices from (26,28) that are available for MAPSS and MP-GRE-WR, allowing for example to improve SNR beyond previous limits (26,28), and also extending these benefits to regular MP-GRE sequences. Finally, by properly choosing small recovery times and large VPS, one can obtain faster $T_{1\rho}$ mapping with similar quality measures as longer configurations, accelerating the acquisition even without undersampling.

## 2. METHODS
### 2.1. 3D-$T_{1\rho}$ magnetization prepared gradient echo pulse sequence

The 3D-$T_{1\rho}$-weighted datasets are usually acquired with various TSLs. In this work, we will follow (33), targeting $T_{1\rho}$ of the human knee joint as an example, by using only two TSLs, of 1ms and 35ms, with spin-lock frequency=500Hz. The pulse sequence includes $T_{1\rho}$ magnetization-preparation, followed by a FLASH sequence (27). The 3D volumes are acquired with Cartesian sampling with two phase-encoding directions, the $k_y$-$k_z$ plane, and one linear readout (frequency-encoding direction), denominated as $k_x$. The $k_y$-$k_z$ plane may be undersampled.

#### 2.1.1. $T_{1\rho}$ Magnetization Preparation

In the three sequences discussed here, the magnetization preparation (MP) step has some differences. As shown in figures 1(a) and 1(b), for MAPSS and MP-GRE-WR, the MP step starts with resetting the longitudinal magnetization (Mz) by using an Mz reset pulse (28,36). The purpose of the Mz reset pulse is to eliminate previous longitudinal magnetization and start signal recovery all from the same initial condition at each segment. The next step is the Mz recovery (usually 300~3000ms), which is essentially a recovery time for the longitudinal magnetization. Then a spin-lock pulse (37), with specified TSL, is applied to generate $T_{1\rho}$ contrast.

Note that the regular MP-GRE does not have an Mz reset pulse (see Figure 1(c)). Compared to the other two variants, this sequence relies on SE reaching a periodic steady state condition, where the SE of each segment is essentially repeated. Due to this, the signal is not acquired in some initial segments, called dummy segments, which are in the transient state.

In MAPSS, there are phase cycling segments, as shown in Figure 1(a), where the spin-lock pulse is modified. The modification is equivalent to a 180º pulse after a standard spin-lock pulse, resulting in rotating the prepared Mz to the negative axis, in the opposite direction when compared to a regular segment.

#### 2.1.2. Gradient Recalled Echo Train

This part is similar in all three sequences, where a FLASH sequence is used, and each acquired echo corresponds to one k-space line of the 3D volume. Standard MP-GRE usually uses constant flip angles (CFA), the same for all TSLs. In MAPSS (28) (see Figure 1(a)), a variable FA (VFA) train is used to reduce the FEs, usually with increased FA for each pulse in the train, with the last FA at 90º. The idea of using VFA train was extended in (26) for MP-GRE sequences with Mz reset (MP-GRE-WR), as shown in Figure 1(b). However, in this pulse sequence, the VFA train is different for each TSL. This is necessary for MP-GRE-WR because the SE of this sequence does not depend only on the initial state, as in MAPSS. In (26), the pulse train for the first TSL is increasing as in MAPSS, optimized for maximum flatness of the SE for predefined relaxation parameters (e.g. $T_1$ and $T_{1\rho}$). But for the following TSLs, the FA in the train may be decreasing to obtain similar levels of flatness.

Until now, there was no optimized VFA for regular MP-GRE sequences. One of the reasons may be because the lack of the Mz reset pulse makes SE too complex to use shot-independent optimization schemes as proposed in (26,28).

### 2.1.3. Acquisition Time

As seen in Figure 1, the acquisition time for each TSL for the last two MP-GRE sequences is given by:

$$T_{tot} = T_{TSLs} \times (D + S) \times (T_{reset} + T_{rec} + TSL + \tau + TR \times VPS) \qquad (1)$$

Where $D$ is the number of dummy shots, $S$ is the number of acquisition shots, usually $S = [N/VPS]$, where $N$ is the number of acquired k-space lines in the $k_y$-$k_z$ plane. In the fully sampled case, $N$ is the total number of phase encoding positions. $T_{TSLs}$ is the number of TSLs Other components of the scan time are described in Figure 1. The $T_{tot}$ for MAPSS is twice what is in Equation 1.

### 2.1.4. k-Space Ordering of Data Collection

The $T_{1\rho}$ contrast is better right after the end of MP. This means that the first echoes have more of the desired contrast, while the later echoes are more contaminated by the changes in SE, showing more FE. Following (26,28), we use a 3D Cartesian center-out approach. We used the one proposed in (38) because it can handle both fully sampled as well as undersampled acquisitions [see figure 1(d) and 1(e)]. This approach, combined with SEs, generates FE in the images that can be of low-pass filtering or high-frequency amplification. This is important because the central area of the k-space is minimally affected by filtering.

The VPS is a choice of the user. Larger VPS means fewer segments to acquire all the desired phase encoding positions, resulting in a faster MRI scan, but probably more FE since the SE will be longer. In this sense, the user must select sequence parameters to balance: SNR, FE, $T_1$ contamination, and scan time.

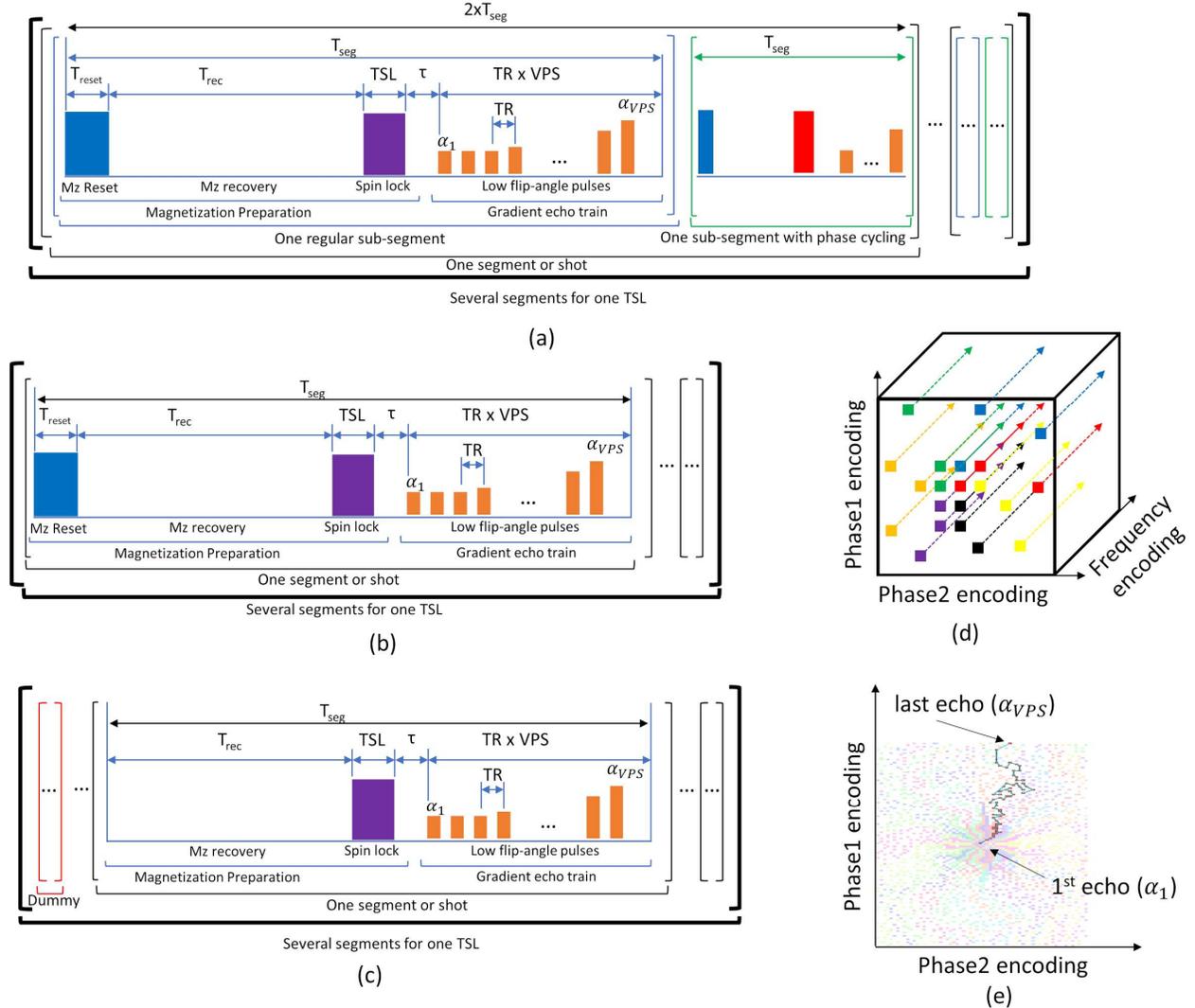

**Figure 1:** Pulse sequences used for $T_{1\rho}$ mapping in this study. (a) The MAPSS pulse sequence uses Mz reset and requires two sub-segments, one regular and one with phase cycling. (b) The MP-GRE with Mz reset (MP-GRE-WR) does not use phase cycling. (c) The regular MP-GRE does not use Mz reset or phase cycling. Because it lacks Mz reset, the SE must achieve steady-state, and some dummy segments may be necessary. (d) The acquisition is Cartesian 3D, at each segment several phase encoding positions given by VPS are acquired, according to the specified SP and ordering scheme. (e) Here a balanced center-out ordering is used, as shown.

## 2.2. Equations of the Signal Evolutions

In the first two sequences, where Mz reset is used, the SE can be described as (39):

$$M_{xy}(n) = A(n)M_{prep} + B(n), \quad (2)$$

where $1 \leq n \leq VPS$, and:

$$A(n) = e_\tau \left[\overbrace{\prod_{i=1}^{n-1} e_1\cos(\alpha_i)}^{1 \text{ if } n=1}\right] e_2 \sin(\alpha_n) \quad (3)$$

$$B(n) = M_0 \left\{(1-e_\tau)\left[\overbrace{\prod_{i=1}^{n-1} e_1\cos(\alpha_i)}^{1 \text{ if } n=1}\right] + (1-e_1)\left[1 + \overbrace{\sum_{p=2}^{n-1}\left(\prod_{i=p}^{n-1} e_1\cos(\alpha_i)\right)}^{\substack{0 \text{ if } n=1 \\ 1 \text{ if } n=2}}\right]\right\} e_2 \sin(\alpha_n) \quad (4)$$

Where $e_\tau = e^{-\frac{\tau}{T_1}}$, $e_1 = e^{-\frac{TR}{T_1}}$, and $e_2 = e^{-\frac{TE}{T_2}}$, The SE is the same for each segment, and:

$$M_{prep} = M_0\left(1 - e^{-\frac{Trec}{T_1}}\right) e^{-\frac{TSL}{T_{1\rho}}} \quad (5)$$

In MAPSS, during the phase cycling segment, the SE is:

$$M_{xy}^{pc}(n) = -A(n)M_{prep} + B(n). \quad (6)$$

The signal from Equation 6 is subtracted from the signal in Equation 2, resulting in $M_{xy}(n) = 2A(n)M_{prep}$. Note that MAPSS signal is free from contamination caused by $B(n)$.

In the third sequence, the regular MP-GRE, without Mz reset, the SEs of different segments are connected:

$$M_{xy}(s,n) = A(n)M_{prep}(s) + B(n), \quad (7)$$

where $s$ represents the segment number:

$$M_{prep}(s) = \left[M_z(s-1, VPS)e^{-\frac{Trec}{T_1}} + M_0\left(1 - e^{-\frac{Trec}{T_1}}\right)\right]e^{-\frac{TSL}{T_{1\rho}}}, \quad (8)$$

and:

$$M_z(s,n) = C(n)M_{prep}(s) + D(n), \quad (9)$$

where $1 \leq n \leq VPS$, and:

$$C(n) = e_\tau \left[ \prod_{i=1}^{n} e_1 \cos(\alpha_i) \right],$$

(10)

$$D(n) = M_0 \left\{ (1 - e_\tau) \left[ \prod_{i=1}^{n} e_1 \cos(\alpha_i) \right] + (1 - e_1) \left[ 1 + \sum_{p=2}^{n} \overbrace{\left( \prod_{i=p}^{n} e_1 \cos(\alpha_i) \right)}^{1 \text{ if } n=1} \right] \right\},$$

(11)

being $M_{prep}(1) = M_0 e^{-\frac{TSL}{T_{1\rho}}}$.

Figure 2 illustrates the SE of 3 shots with VPS=128, $TR$=6ms, $TE$=3ms, $\tau$=3ms, and $Trec$=1110ms for each sequence considering three different materials with different relaxation values, assuming $M_0 = 1$. Note that because the SE in the third sequence is coupled between segments, it is more complex to model, and consequently, to optimize. The MAPSS (28) and MP-GRE-WR (26) already include VFA with optimization for maximum flatness for the component $(T_1, T_2, T_{1\rho})$=(1200ms, 35ms, 40ms), following their references. The MP-GRE here uses a constant flip-angle of 9 deg (MP-GRE-CFA). Note the optimization of MAPSS and MP-GRE-WR can be carried out using only one segment, while the optimization of MP-GRE (without Mz reset) needs to include all segments, including the dummy ones.

Note that SE also depends on the relaxation parameters, such as $T_1$, $T_2$, and $T_{1\rho}$. In the optimization process, we may be interested in optimal performance on several different sets of relaxation values. We will use $M_{xy}(k, t, s, n)$ to denote the SE with relaxation sets $1 \leq k \leq K$, where $K$ is the number of relaxation sets, being $[T_1(k), T_2(k), T_{1\rho}(k)]$, for the $1 \leq t \leq T_{TSLs}$, where $T_{TSLs}$ is the number of TSLs, on the segment $1 \leq s \leq S + D$, after the flip-angle pulse $1 \leq n \leq VPS$.

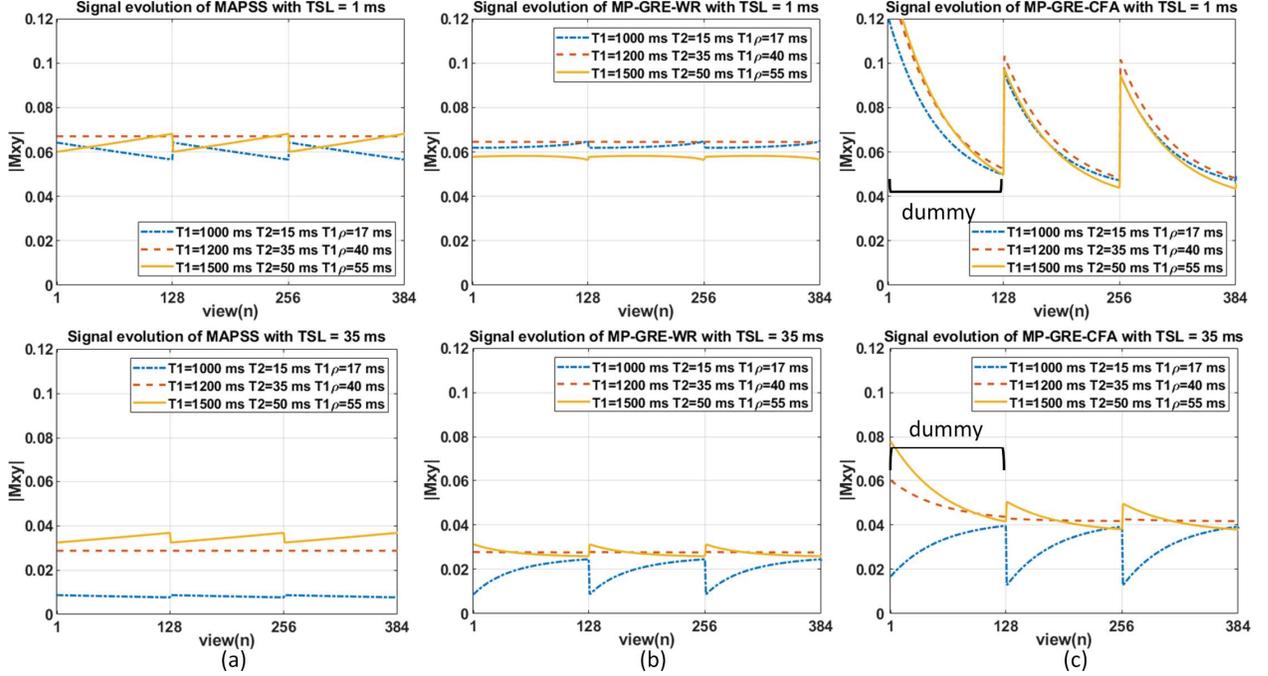

**Figure 2:** SE for the three different sequences for TSL=1ms and TSL=35ms for three different components with different relaxation parameters. (a) SE of MAPSS, after subtracting phase cycling segments. (b) SE of MP-GRE-WR (with Mz reset). (c) SE of standard MP-GRE sequences (without Mz reset) with constant FA. Note that in (c), the first D dummy segments in MP-GRE have a different SE from the other S (acquired) segments.

### 2.3. Proposed Optimization of the Variable Flip-Angles (OVFA)

The optimization tries to find the best VFA for different TSLs that maximize a surrogate function of the precision and accuracy in $T_{1\rho}$ mapping. More precisely, we maximize a function composed of three SE terms and one constraint-like term that deal with different imperfections in the SE of the $T_{1\rho}$ mapping sequences. The general form of the cost function is:

$$\widehat{\boldsymbol{\alpha}} = \underset{\boldsymbol{\alpha}}{\mathrm{argmax}} \left[ \sum_{k=1}^{K} (A(k, \boldsymbol{\alpha}) + F(k, \boldsymbol{\alpha}) + S(k, \boldsymbol{\alpha})) \right] + R(\boldsymbol{\alpha}),$$

(12)

where $k$ is the index of the relaxation set of interest and $\boldsymbol{\alpha}$ is a vector of size $T_{TSLs} \cdot VPS \times 1$, containing the flip angles train for all the TSLs. The flip angles train may be the same for all TSLs as in MAPSS, or it may be different for each TSL as in MP-GRE-WR and MP-GRE.

The first term, $A(k, \boldsymbol{\alpha})$, is chosen to improve *accuracy*, particularly at the beginning of each segment of the SE where the components are related to the acquisition of the central area of the k-space. This term counters the $T_1$ contamination caused by the delay $\tau$ in the pulse sequence as well as the differences in $M_{prep}(s)$ in the MP-GRE sequence.

In this work, we used:

$$A(k, \alpha) = -\lambda_A \|Am_k(\alpha)\|_2^2, \tag{13}$$

where $\boldsymbol{m}_k(\boldsymbol{\alpha})$ is the *normalized* SE, $\boldsymbol{m}_k(\boldsymbol{\alpha}) = [M_{xy}(k, t_1, 1, 1)/e^{-\frac{t_1}{T1\rho(k)}} \cdots M_{xy}(k, t_{T_{TSLs}}, S + D, VPS)/e^{-\frac{t_{T_{TSLs}}}{T1\rho(k)}}]$, of the $k$-th component of the relaxation parameter set, stacked in a vector for all TSLs, segments, and pulses, with size $T_{TSLs} \cdot (S + D) \cdot VPS \times 1$. The SE is normalized by the effect of the $T1\rho(k)$ relaxation, $e^{-\frac{t}{T1\rho(k)}}$, which is different for each TSL. The matrix $\boldsymbol{A}$ computes the finite difference between all pairs of $M_{xy}(k, t_p, s, 1)/e^{-\frac{t_p}{T1\rho(k)}}$ and $M_{xy}(k, t_q, s, 1)/e^{-\frac{t_q}{T1\rho(k)}}$, being $t_p$ and $t_q$ two different TSLs. This is done for the first elements for all acquired segments $s$ (except the dummy segments). The idea behind this term is that if this difference is minimized, the accuracy of $T_{1\rho}$ mapping is improved in the low-frequency components. Because the SE of these components should differ only due to the normalization factor $e^{-\frac{t}{T1\rho(k)}}$. Due to this term, initial $\alpha_n$ values may be slightly different for each TSL. The $\lambda_A$ is the weighting factor of this term of optimization. As we will discuss later, this term is not necessary for the optimization of MAPSS sequence because it uses the same $\boldsymbol{\alpha}$ for all TSLs. Also, in MAPSS the $T_1$ contamination effects caused by the delay $\tau$ are solved by using phase cycling acquisitions.

The second term, $\boldsymbol{F}(k, \boldsymbol{\alpha})$, is designed to improve the *flatness* of the SE, consequently making the modulation transfer function (MTF) (29), which represents the k-space filtering, as constant as possible. This term counters the FE caused by the change in the magnitude of the measured signal due to $T_1$ relaxation that happens during SE within a segment. In this work, we used:

$$F(k, \alpha) = -\lambda_F \|Fm_k(\alpha)\|_2^2, \tag{14}$$

Where $\boldsymbol{m}_k(\boldsymbol{\alpha})$ is the same as before. The matrix $\boldsymbol{F}$ computes the finite difference on the SE inside the segment, for $1 \leq n \leq VPS$, and it is repeated for all TSLs and all $s$ acquired segments (except dummy segments). This term enforces VFA to remove FE as in MAPSS (28) and tailored VFA (26) approaches. The $\lambda_F$ is the weighting factor of this term of the optimization.

The third term, $\boldsymbol{S}(k, \boldsymbol{\alpha})$, is designed to improve the *signal strength* at the beginning of the SE. This term focus on improving SNR, by forcing the components of the SE related to low-frequency k-space components to be close to the desired signal intensity. In this work, we used:

$$S(k, \boldsymbol{\alpha}) = -\lambda_S \|S(\boldsymbol{m}_k(\boldsymbol{\alpha}) - \boldsymbol{m}_{ref})\|_2^2, \tag{15}$$

where $\boldsymbol{m}_k(\boldsymbol{\alpha})$ is the same as before, $\boldsymbol{m}_{ref}$ is the reference signal, and the matrix $S$ has ones in the positions we want $\boldsymbol{m}_k(\boldsymbol{\alpha})$ to be close to $\boldsymbol{m}_{ref}$, and zeros on the others. In this work, $S$ has ones on the positions related to $M_{xy}(t_1, k, s, 1)/e^{-\frac{t_1}{T1\rho(k)}}$, the first element of the first TSL, for all segments, except the dummy segments. We found that it is not needed to force this on the other TSLs because Equation 13 does this job. The $\lambda_S$ is the weighting factor of this term of optimization.

Optionally, one may consider specific constraints to apply desired behavior in the flip-angles, such as in MAPSS where the last FA should be 90º.

$$R(\boldsymbol{\alpha}) = -\lambda_R \|R(\boldsymbol{\alpha} - \boldsymbol{\alpha}_{ref})\|_2^2. \tag{16}$$

The matrix $R$ has ones to force the elements of $\boldsymbol{\alpha}$ we want to be close or equal to the reference, letting the others be defined by the optimization. The $\lambda_R$ is the weighting factor of this term of optimization.

We can also write Equation 12 as:

$$\widehat{\boldsymbol{\alpha}} = \underset{\boldsymbol{\alpha}}{\operatorname{argmin}} \left[ \sum_{k=1}^{K} \omega_k \left( \lambda_A \|A\boldsymbol{m}_k(\boldsymbol{\alpha})\|_2^2 + \lambda_F \|F\boldsymbol{m}_k(\boldsymbol{\alpha})\|_2^2 + \lambda_S \|S(\boldsymbol{m}_k(\boldsymbol{\alpha}) - \boldsymbol{m}_{ref})\|_2^2 \right) \right] \\ + \lambda_R \|R(\boldsymbol{\alpha} - \boldsymbol{\alpha}_{ref})\|_2^2, \tag{17}$$

Where $\omega_k$ is a weight that is given to each particular relaxation parameter according to its importance in the set. In this work we used $\omega_k = |T1\rho(k)|^2 / \sum_{i=1}^{K} |T1\rho(i)|^2$, since it produced results with more uniform errors across different sets of parameters.

Ideally, the optimization searches for flip-angles $\boldsymbol{\alpha}$ that produce normalized SEs, $\boldsymbol{m}_k(\boldsymbol{\alpha})$, as constant as possible and close to $\boldsymbol{m}_{ref}$, prioritizing the desired properties of the SE related to low-frequency k-space components. If that happens the SE will have all undesirable factors that affect accuracy, such as $T_1$ contamination and FE, removed or minimized, while $T_{1\rho}$ contrast will be preserved. At the same time, the signal should have the desired strength, defined by $\boldsymbol{m}_{ref}$, that can be chosen to improve SNR and, consequently, the precision of $T_{1\rho}$ mapping. Equation 17 is a non-linear least squares problem and was solved using the trust region conjugate gradient (TRCG) method (33).

### 2.4. Synthetic Profile Evaluation

We evaluated the theoretical performance of the optimized VFA using simulated signals, as seen in Figure 3(a), assuming a profile as shown in Figure 3(c). The ideal profile undergoes signal modulation in the frequency domain (Fig. 3(b)), according to the SE of the specific pulse sequence with all the relaxation parameters tested. This will apply the desired $T_{1\rho}$ contrast, but will also apply imperfections of the SE, such as $T_1$ contamination and FE. White Gaussian noise, with a standard deviation equivalent to the estimated from real acquisitions, is also added to simulate the noise effects to evaluate SNR improvements, as shown in Figure 3(e). The $T_{1\rho}$ mapping is evaluated using artificially distorted profiles, as shown in Figure 3(f)-(g).

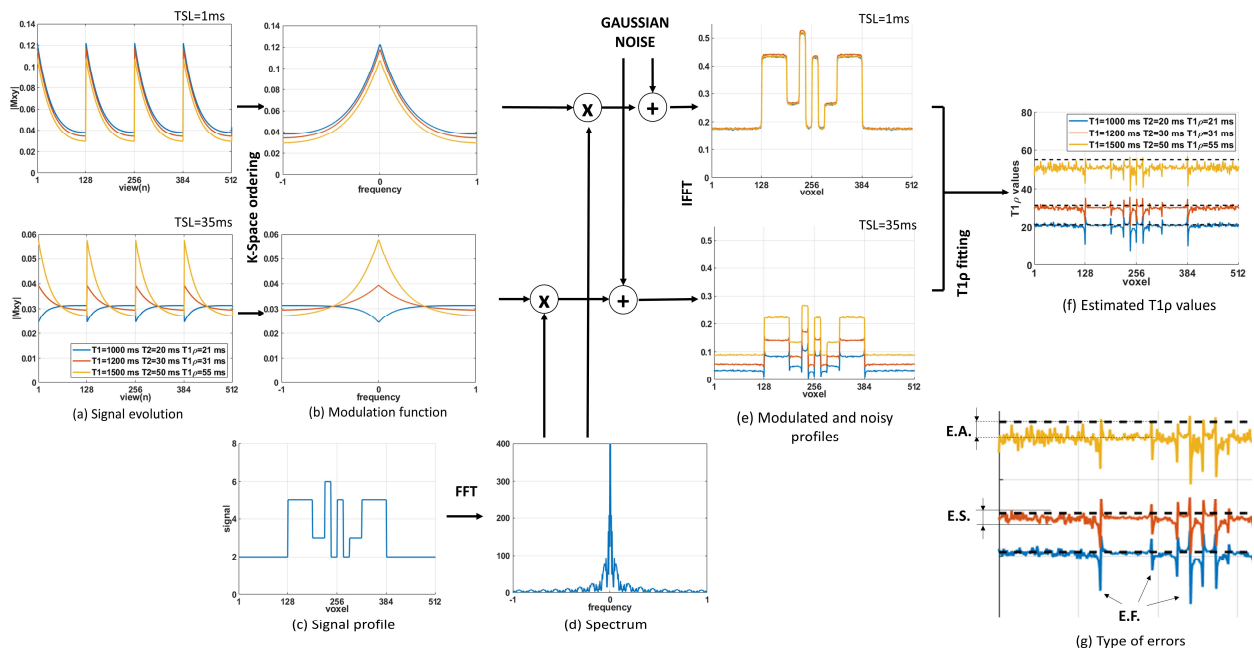

**Figure 3:** Simulated evaluation of the sequences using the effects of the SE on the idealized profile, as shown in (c), with spectrum as shown in (d). The SEs are different for different TSLs and relaxation parameters, as shown in (a), leading to different modulation transfer functions, as shown in (b). The resulting modulated and noisy profiles are shown in (e). After $T_{1\rho}$ fitting, as shown in (f), three types of errors are seen in the estimated $T_{1\rho}$ values, as shown in (g). The E.A. is the error in accuracy, usually related to the error in the first echoes of the echo train, related to the central k-space; the E.F is the error due to the FE, which is usually larger in the discontinuities of the profile; and E.S is the error due to noise, which is larger when the SNR is small.

Note that three kinds of errors are noted in the $T_{1\rho}$ values: E.A. is inaccuracies due to $T_1$ contamination that affects the first echoes, close to the center of the k-space, biasing the estimated values (visible as a shift of the estimated value from the measurements); E.F. is the error due to FE, due to the lack of flatness in the sequence (visible as peaks of large error that happen around the edges of the given profile); and E.S. that is the error due to noise, due to the limited signal strength produced by the sequence (visible as a random oscillation of the measured signal around the mean value). In the optimization of Equation 12, E.A. is countered by the first term (Equation 13), E.F. is reduced by the second term (Equation 14), and E.S. is reduced by the third term (Equation 15). These errors are different for each set of relaxation values.

## 2.5. Image Reconstruction, Fitting, and Evaluation

The image reconstruction is performed when k-space data is acquired from the MRI scanner. The 3D volume for each TSL is separated into 2D slices by applying FFT in the readout direction, each slice is reconstructed with SENSE (40), which minimizes:

$$\hat{x}_t = \underset{x_t}{\mathrm{argmin}} \|y_t - EBx_t\|_2^2, \qquad (18)$$

where $x_t$ is a complex-valued vector that represents an image with TSL $t$, with size $N_y \times N_z$, with $N_y$ being the image size in the y-axis and $N_z$ the size in the z-axis. The vector $y_t$ represents the captured k-space with size $N_y \times N_z \times N_c$, where $N_c = 15$ is the number of coils. The matrix $B$ contains the coil sensitivities and phase compensation (15,41), $E$ the Fourier transforms of all sensitivity-weighted images. The $\|e\|_2^2$ is the squared $l_2$-norm or Euclidean norm of $e = y_t - EBx_t$.

The $T_{1\rho}$ fitting is performed using monoexponential models given by:

$$f(t, n) = a(n) \, exp\left(-\frac{t}{p(n)}\right), \qquad (19)$$

Where $a(n)$ is the complex-valued amplitude at the voxel $n$ and $p(n)$ corresponds to the $T_{1\rho}$ values. For the fitting, we used a shallow neural network (NN) (42), previously used in (33). This NN performed similarly to non-linear least squared fitting (43), the details of the architecture and training of the NN can be found in (33). To estimate the standard deviation of the noise and the respective SNR or acquired images, we use the Marchenko–Pastur Principal Component Analysis (MP-PCA) method from (44).

The quality of $T_{1\rho}$ mapping is evaluated in terms of the mean of normalized absolute difference (MNAD), denoted as:

$$MNAD(ROI) = \underset{n \in ROI}{\text{mean}} \left( \frac{|p(n) - p_{ref}(n)|}{|p_{ref}(n)|} \right), \qquad (21)$$

In the simulated performance tests, the reference, $p_{ref}$, are the ground truth values. For model agarose phantoms and healthy volunteers the ground truth is unknown, then the best method in the simulated experiments, MAPSS with optimized VFA (MAPSS-OVFA), was used as a reference. We named this evaluation Ref-MNAD. To evaluate the effects of noise only, the methods were compared against a repetition of themselves, results were named Repeat-MNAD. We also computed the histograms to observe the distribution of the measured $T_{1\rho}$ values.

Note that the MTFs cause FE, which is different for each TSL and each different relaxation parameter set. We computed the predicted FWHM of their corresponding point spread functions (45) as a way to evaluate numerically (shown in voxel numbers) the loss of detail caused by the FE.

## 2.6. Compared Sequences

We applied the optimization of the variable flip-angles with 3 different sequences: MAPSS, MP-GRE-WR, and regular MP-GRE. We use OVFA with two objectives: 1) improving the precision by improving SNR and 2) making the sequence faster by using a different setup. The faster setup uses a smaller $T_1$ recovery time (Trec) and higher VPS, which usually leads to a loss of quality. In this case, OVFA should minimize these problems.

The MP-GRE is the only sequence that can use Trec equal to zero. This new configuration, denoted MP-GRE OVFA ZRT, is only achievable because MP-GRE does not use Mz reset pulse, so it can exploit the remaining Mz of previous segments. In this case, we use OVFA for three different purposes: 1) to improve SNR, 2) to improve flatness (26,28), and 3) to improve speed, using an undersampling factor (UF) of 3. This sequence uses an undersampling pattern learned with BASS (18) for SENSE reconstructions.

We also use the MP-GRE OVFA ZRT to obtain $T_{1\rho}$ mapping with improved spatial resolution, using four times the in-plane resolution (2x horizontal and 2x vertical). This sequence is denominated MP-GRE OVFA ZRT HR, and it is used to illustrate the potential of optimized VFA for high-resolution $T_{1\rho}$ mapping.

### 2.6.1. Parameters for the Optimization of the VFA

We assume our target object is composed of 9 sets of relaxation parameters. They are generated by a 3x3 grid of $T_1$ relaxation (900, 1200, and 2000 ms) and $T_{1\rho}$ relaxation (13, 32, 55 ms). We assume the following T₂ values (11, 30, 50 ms), that are paired with the $T_{1\rho}$ values shown. The weighting parameters ($\lambda_A, \lambda_F, \lambda_S, \lambda_R$) used in the optimization were systematically selected to minimize the predicted MNAD in the simulated case. The values of $\boldsymbol{m_{ref}}$ were also chosen with the same purpose, except in the FAST methods, which were chosen to achieve nearly the same initial magnitude of SE as their standard versions. The used values are shown in Table 1.

**Table 1:** Compared different pulse sequences with their weighting parameters ($\lambda_A, \lambda_F, \lambda_S, \lambda_R$), $\boldsymbol{m_{ref}}$ and $\boldsymbol{\alpha_{ref}}$ values used for optimization.

| Sequence Name | Optimization purpose | $\lambda_A$ | $\lambda_F$ | $\lambda_S$ | $\lambda_R$ | $\boldsymbol{m_{ref}}$ | $\boldsymbol{\alpha_{ref}}$ |
|---|---|---|---|---|---|---|---|
| MAPSS | Flatness | Uses (28) | | | | | |
| MAPSS-OVFA | SNR | 0 | 2 | 100 | 100 | 0.145 | 90° |
| *FAST* MAPSS-OVFA | Speed | 0 | 100 | 100 | 100 | 0.075 | 90° |
| MP-GRE-WR | Flatness | Uses (26) | | | | | |
| MP-GRE-WR-OVFA | SNR and Accuracy | 1000 | 10 | 10 | 0 | 0.13 | n/a |
| *FAST* MP-GRE-WR-OVFA | Speed and Accuracy | 100 | 10 | 10 | 0 | 0.08 | n/a |
| MP-GRE CFA | Non-optimized | Constant FA | | | | | |
| MP-GRE-OVFA | SNR and Accuracy | 1000 | 5 | 10 | 0 | 0.14 | n/a |
| *FAST* MP-GRE-OVFA | Speed and Accuracy | 1000 | 10 | 10 | 0 | 0.11 | n/a |
| MP-GRE-OVFA ZRT FLAT | Flatness and Accuracy | 3000 | 150 | 10 | 0 | 0.065 | n/a |
| MP-GRE-OVFA ZRT SNR+ | SNR and Accuracy | 3000 | 50 | 10 | 0 | 0.095 | n/a |
| MP-GRE-OVFA ZRT FAST | Speed and Accuracy | 3000 | 100 | 10 | 0 | 0.08 | n/a |
| MP-GRE-OVFA ZRT HR | SNR and Accuracy | 3000 | 100 | 10 | 0 | 0.08 | n/a |

### 2.7. MRI Scanning Specifications

The data was acquired on a Prisma 3T scanner (Siemens Healthcare Gmbh, Erlangen, Germany) with a vendor-provided 1-Tx/15-Rx knee coil (QED, OH). The 3D volume has 64 slices with a data matrix size of 256×256 voxels (HR uses 512×512 voxels). The linear readout (frequency encoding) direction acquires 256 samples (HR uses 512 samples), with N=256×64 phase encoding positions in the fully sampled case (HR uses N=512×64). The number of segments (see Fig 1) to acquire all data at each TSL is shown in Table 1 for each sequence. We use binomial pulses for water excitation only (46,47). Also, the number of shots (S), dummy shots (D), VPS, and Trec are included in Table 2, and $\tau = TR/2$.

We acquire two TSLs of 1 ms and 35 ms, following (33). The MAPSS sequences always use double the total acquisition time for the same configuration, due to phase-cycled acquisitions (28).

### 2.8. $T_{1\rho}$ in Model Agarose Phantoms and Human Knee Joint:

We use agarose tubes with concentrations 3%, 4%, 5%, 6%, and 8% as phantoms. The MRI scans have a voxel size of 0.8mm×0.8mm×2mm (HR uses 0.4mm×0.4mm×2mm), with FOV of 200mm×200mm×128mm. We repeated the scans to evaluate repeatability with phantoms. We scanned five healthy volunteers, with a mean age of 35 and SD=6.4 years old. The MRI scans have a voxel size of 0.7mm×0.7mm×2mm (HR uses 0.35mm×0.35mm×2mm), with FOV of 180mm×180mm×128mm. This study was approved by the institutional review board (IRB) of New York University Langone Health and was compliant with the health insurance portability and accountability act (HIPAA). All volunteers provided their consent before MRI scanning.

## 3. RESULTS

The numerical results of all different sequences are shown in Table 2. Note that all three sequences improved SNR and reduced MNAD when OVFA was used to improve SNR. The faster versions used different parameters that halve the total acquisition time, usually causing degradation in the performance. However, by using OVFA the faster versions achieved nearly the same quality as the original versions of the sequences. This illustrates that OVFA can make a sequence faster and/or more precise.

The precision was measured by the repeat-MNAD. The accuracy of the sequences (systematic errors) can be indirectly observed by the difference between Ref-MNAD and Repeat-MNAD. Note that MAPSS sequences are the most accurate of all three since Ref-MNAD and Repeat-MNAD are virtually the same. However, the proportion of systematic error in MP-GRE and MP-GRE-WR was reduced with OVFA, which indicates that these sequences were also made more accurate with OVFA. Repeat-MNAD was not measured with human volunteers to avoid long scanning sessions.

The FE, described by the MTF, caused a small loss of detail, the FWHM of the point spread functions of each MTF for each relaxation parameter set was measured and their average is shown in Table 2, for the first and the second TSL, respectively. MAPSS has the same FEs for all TSLs, but MP-GRE and MP-GRE-WR have more FE in the first TSL. The high-frequency amplification that usually happens with long TSLs was very minor in our experiments. The sequences optimized for maximum flatness obtained the least FE.

**Table 2:** Tested sequences, parameters, and results. Common parameters are 2 TSLs (1ms, 35ms), and TR of 6ms (except MP-GRE-OVFA ZRT HR that uses TR of 8.6 ms). For each sequence, it is specified views per segment (VPS), the number of acquired segments or shots (S), $T_1$ recovery time ($T_{rec}$), dummy shots (D), the time for each TSL (Time), which does not include phase cycling time, and the total time of the acquisition (Total time), which include all TSLs and phase cycling acquired.

| Sequence FA | Parameters for each TSL or PC | | | | | Total time (min) | FW HM | Evaluation | Sim. Phan. | Phan. | Sim. Vol. | Vol. |
|---|---|---|---|---|---|---|---|---|---|---|---|---|
| | VPS | S | Trec (ms) | D | Time (min) | | | | | | | |
| MAPSS | 128 | 128 | 1110 | 0 | 4 | 16 | 1.0 1.0 | SNR | 12.7 | 12.4 | 5.0 | 5.5 |
| | | | | | | | | Ref-MNAD | 7.6% | 6.5% | 22% | 20% |
| | | | | | | | | Repeat-MNAD | 7.6% | 6.5% | 22% | - |
| MAPSS-OVFA | 128 | 128 | 1110 | 0 | 4 | 16 | 1.4 1.4 | SNR | **24.0** | **23.4** | **10.1** | **11.0** |
| | | | | | | | | Ref-MNAD | **4.0%** | **4.0%** | **12%** | **12%** |
| | | | | | | | | Repeat-MNAD | **4.0%** | **4.0%** | **12%** | - |
| *FAST* MAPSS-OVFA | 256 | 64 | 340 | 0 | **2** | **8** | 3.1 3.1 | SNR | 13.8 | 9.7 | 5.5 | 5.9 |
| | | | | | | | | Ref-MNAD | 7.9% | 7.6% | 18% | 17% |
| | | | | | | | | Repeat-MNAD | 7.9% | 7.6% | 18% | - |
| MP-GRE-WR | 128 | 128 | 1110 | 0 | 4 | 8 | 1.0 1.0 | SNR | 14.3 | 16.9 | 6.6 | 8.2 |
| | | | | | | | | Ref-MNAD | 7.4% | 7.7% | 17% | 17% |
| | | | | | | | | Repeat-MNAD | 6.8% | 6.8% | 16% | - |
| MP-GRE-WR-OVFA | 128 | 128 | 1110 | 0 | 4 | 8 | 1.4 1.0 | SNR | **26.2** | **26.2** | **12.9** | **15.7** |
| | | | | | | | | Ref-MNAD | **4.8%** | **5.1%** | **12%** | **14%** |
| | | | | | | | | Repeat-MNAD | **3.9%** | **4.3%** | **11%** | - |
| *FAST* MP-GRE-WR-OVFA | 256 | 64 | 340 | 0 | **2** | **4** | 1.2 1.0 | SNR | 16.5 | 18.9 | 10.1 | 10.3 |
| | | | | | | | | Ref-MNAD | 8.5% | 8.4% | 17% | 17% |
| | | | | | | | | Repeat-MNAD | 5.9% | 5.3% | 16% | |
| MP-GRE-CFA | 128 | 128 | 1064 | 3 | 4 | 8 | 1.2 1.0 | SNR | 24.2 | 26.6 | 11.9 | 13.0 |
| | | | | | | | | Ref-MNAD | 6.0% | 6.3% | 14% | 14% |
| | | | | | | | | Repeat-MNAD | 3.6% | 3.2% | 12% | - |
| MP-GRE-OVFA | 128 | 128 | 1064 | 3 | 4 | 8 | 1.4 1.1 | SNR | **33.7** | **35.1** | **13.4** | **16.9** |
| | | | | | | | | Ref-MNAD | **4.2%** | **4.5%** | **12%** | **13%** |
| | | | | | | | | Repeat-MNAD | **2.9%** | **2.8%** | **10%** | - |
| *FAST* MP-GRE-OVFA | 256 | 64 | 228 | 4 | **2** | **4** | 1.6 1.1 | SNR | 25.3 | 28.5 | 11.4 | 13.3 |
| | | | | | | | | Ref-MNAD | 5.5% | 5.5% | 14% | 14% |
| | | | | | | | | Repeat-MNAD | 3.6% | 3.5% | 12% | - |
| MP-GRE-OVFA ZRT FLAT | 128 | 128 | 0 | 8 | 1.75 | 3.5 | 1.0 1.0 | SNR | 12.2 | 11.8 | 6.7 | 6.3 |
| | | | | | | | | Ref-MNAD | 10.1% | 9.9% | 17% | 20% |
| | | | | | | | | Repeat-MNAD | 8.1% | 8.8% | 15% | - |
| MP-GRE-OVFA ZRT SNR+ | 128 | 128 | 0 | 8 | 1.75 | 3.5 | 1.4 1.0 | SNR | **18.8** | **17.5** | **9.6** | **9.0** |
| | | | | | | | | Ref-MNAD | **6.8%** | **6.8%** | **15%** | **16%** |
| | | | | | | | | Repeat-MNAD | **5.5%** | **5.9%** | **12%** | - |
| MP-GRE-OVFA ZRT FAST | 43 UF3 | 128 | 0 | 8 | **0.65** | **1.3** | 1.2 1.0 | SNR | 14.2 | 14.4 | 8.9 | 7.7 |
| | | | | | | | | Ref-MNAD | 8.6% | 9.2% | 16% | 17% |
| | | | | | | | | Repeat-MNAD | 6.6% | 7.6 | 14% | - |
| MP-GRE-OVFA ZRT HR | 256 | 128 | 0 | 8 | 4.85 | 9.7 | 1.2 1.0 | SNR | 9.0 | 8.2 | 5.0 | 4.7 |
| | | | | | | | | Ref-MNAD | 12.8% | - | 20% | - |
| | | | | | | | | Repeat-MNAD | 11.7% | 11.6% | 19% | - |

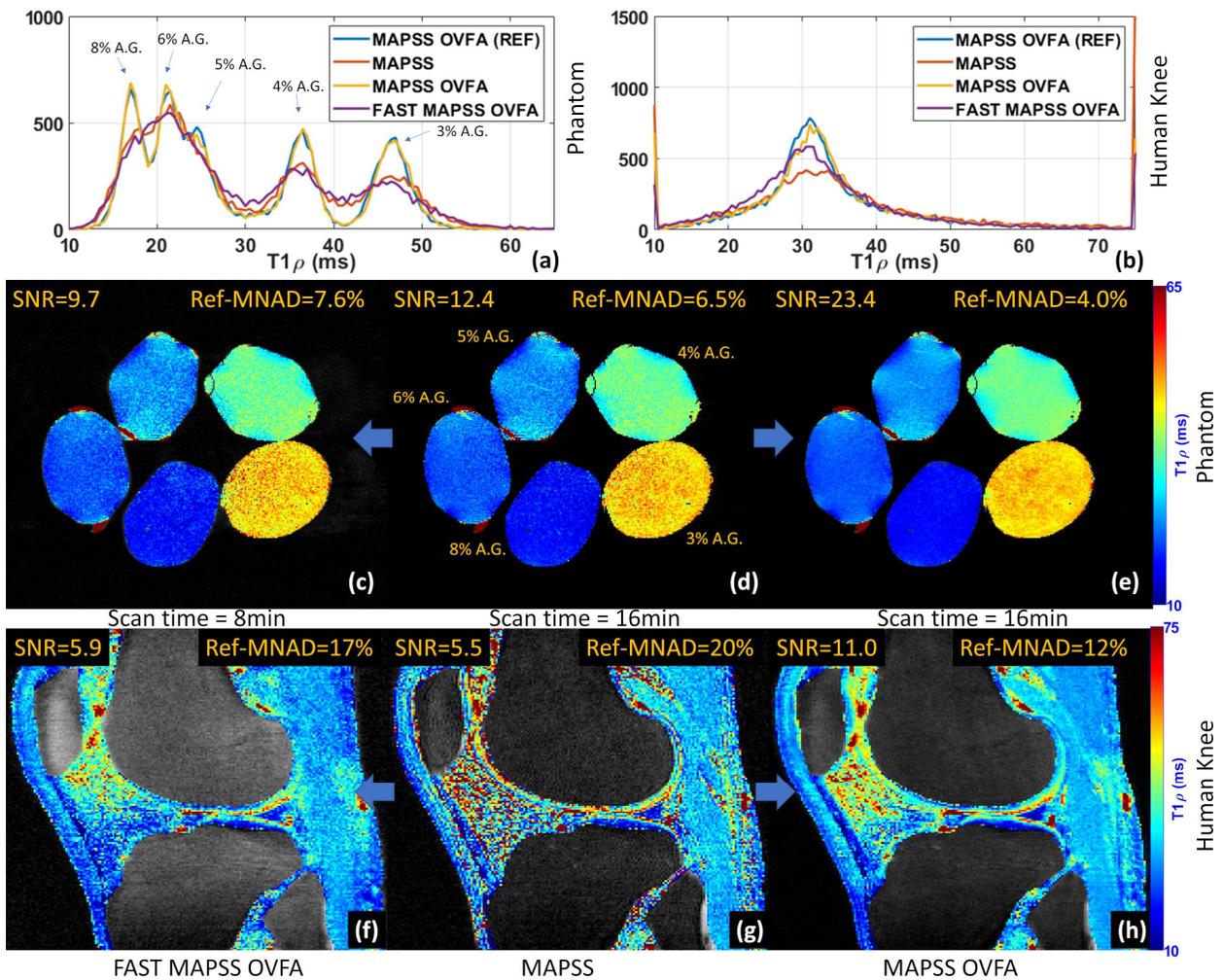

**Figure 4:** Illustration of the performance of the OVFA with the MAPSS sequence. In (a) and (b) the histograms for the model phantom and human knee joint are shown. In (c) and (f) the $T_{1\rho}$ maps of the faster versions are shown for model phantoms and human knee joints. In (d) and (g) the $T_{1\rho}$ maps of the ordinary MAPSS sequence are shown. And in (e) and (h), the $T_{1\rho}$ maps of the MAPSS OVFA are shown for model phantom and human knee joint.

In Figure 4, the SNR improvements in MAPSS-OVFA can be easily seen, compared to MAPSS. The histogram of the agarose phantoms using MAPSS-OVFA clearly shows the 5 peaks related to each tube with concentrations of 3%, 4%, 5%, 6%, and 8%. In the ordinary MAPSS and FAST MAPSS OVFA, the noise spreads the peaks shown in the histograms, for example, the peaks related to 5%, 6%, and 8% agarose are seen as only one wider peak.

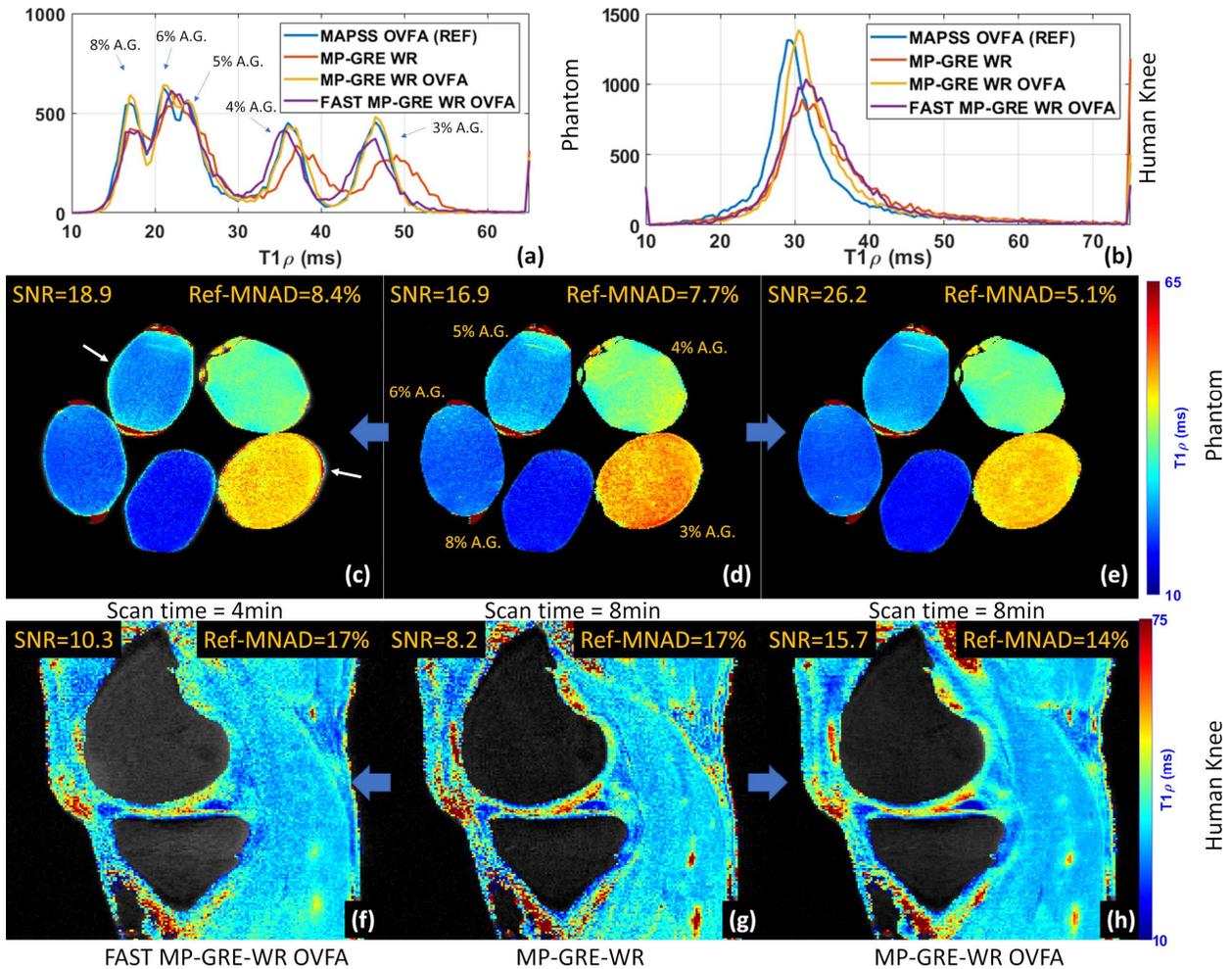

**Figure 5:** Illustration of the performance of the OVFA with the MP-GRE-WR sequence. In (a) and (b) the histograms for phantoms and human knee joints are shown. In (c) and (f) the $T_{1\rho}$ maps of the fast versions are shown for phantoms and human knee joints. In (d) and (g) the $T_{1\rho}$ maps of the ordinary MP-GRE-WR sequence are shown. In (e) and (h), the $T_{1\rho}$ maps of the MP-GRE-WR OVFA are shown for model phantom and human knee joint.

In Figure 5, we can see the SNR and accuracy improvement of OVFA in MP-GRE-WR, when compared to the standard version. The histogram of the agarose phantoms of the MP-GRE-WR-OVFA is much closer to the reference (MAPSS-OVFA). In the ordinary MP-GRE-WR, there is an overestimation of the large $T_{1\rho}$ values, that were fixed with OVFA. The faster version still has some remaining FEs visible (see white arrows).

In Figure 6 we can see improvement in all aspects (SNR, accuracy, and some reduction in FE) with OVFA in MP-GRE when compared to the version with CFA. The histogram of the agarose phantoms of the MP-GRE-OVFA is much closer to the reference (MAPSS-OVFA). In the ordinary

MP-GRE-CFA, there is a small underestimation of the large $T_{1\rho}$ values, which was improved with OVFA. The MP-GRE is the sequence with the best SNR of all. In all cases, the agarose gel peaks are distinct.

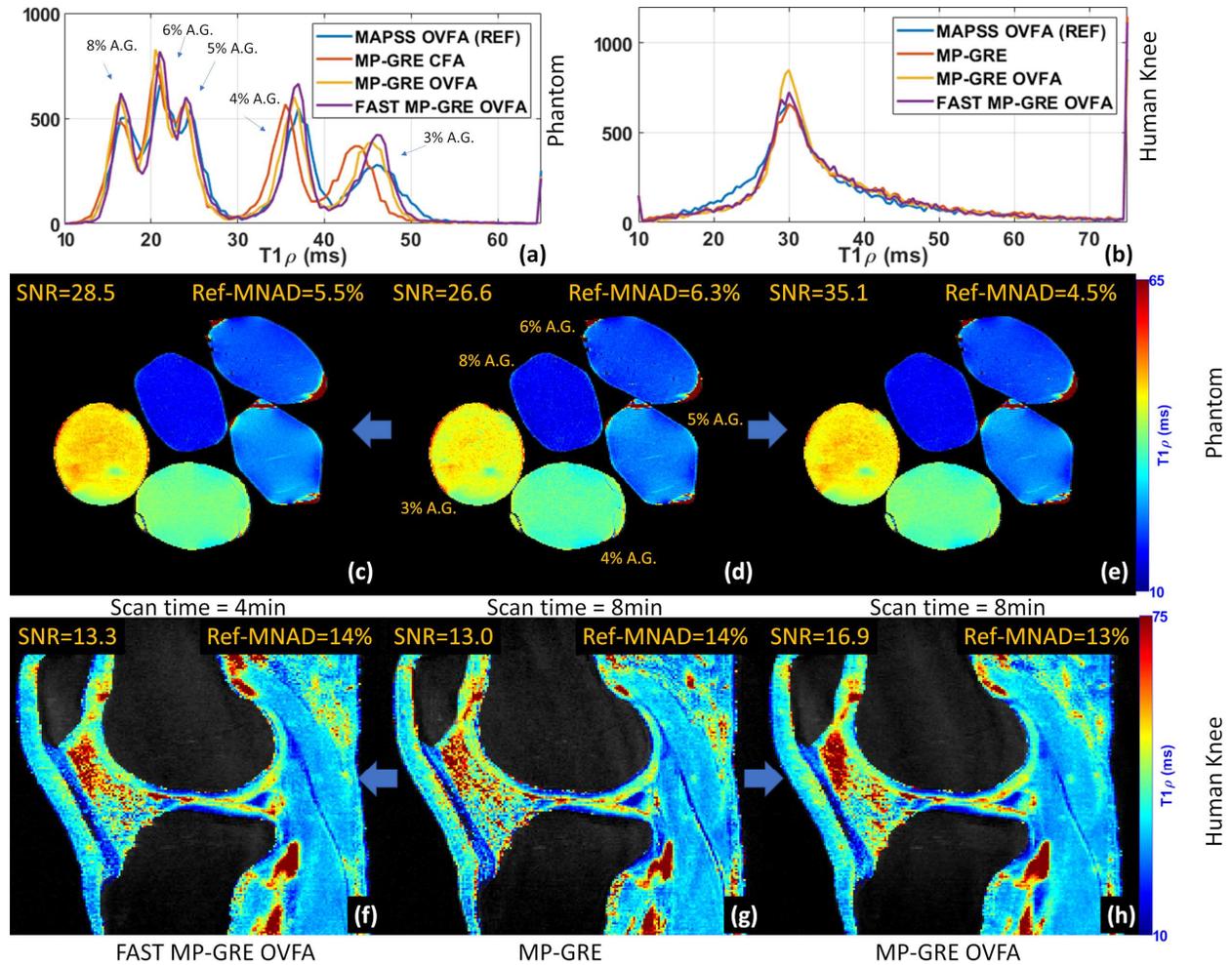

**Figure 6:** Illustration of the performance of the OVFA with the MP-GRE sequence. In (a) and (b) the histograms for model phantoms and human knee joints are shown. In (c) and (f) the $T_{1\rho}$ maps of the faster versions are shown for model phantoms and human knee joints. In (d) and (g) the $T_{1\rho}$ maps of the ordinary MP-GRE CFA sequence are shown. In (e) and (h), the $T_{1\rho}$ maps of the MP-GRE OVFA are shown for model phantoms and human knee joints.

In Figure 7 we can see visually the quality of MP-GRE-OVFA with ZRT when the optimization targets are more SNR, more flatness (fewer FE), or when it is combined with undersampling (UF=3) for more speed. The fast version has more remaining FE visible (see white arrows).

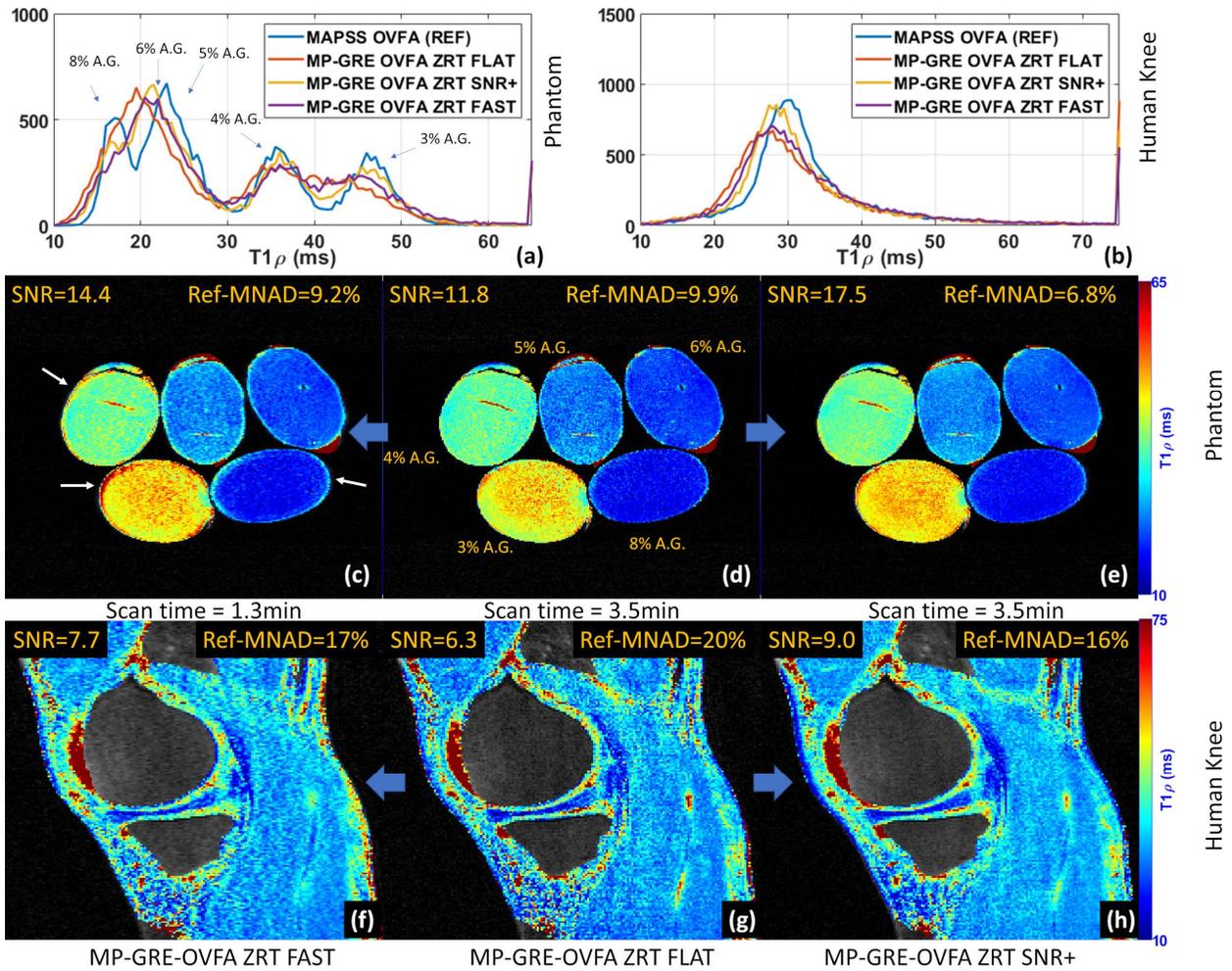

**Figure 7:** Illustration of the performance of the OVFA with the MP-GRE sequence with ZRT. In (a) and (b) the histograms for model phantoms and human knee joints are shown. In (c) and (f) the $T_{1\rho}$ maps of the versions with undersampling are shown for model phantoms and human knee joints. In (d) and (g) the $T_{1\rho}$ maps of the version with more flatness. In (e) and (h), the $T_{1\rho}$ maps of the version with more SNR are shown for model phantoms and human knee joints.

In Figure 8, we illustrate the performance of the MP-GRE-OVFA-ZRT-HR, comparing the HR $T_{1\rho}$ maps with MAPSS-OVFA. The improvement in details because of the better resolution can make $T_{1\rho}$ mapping useful in applications such as cartilage repair (48), which requires investigation of smaller structures.

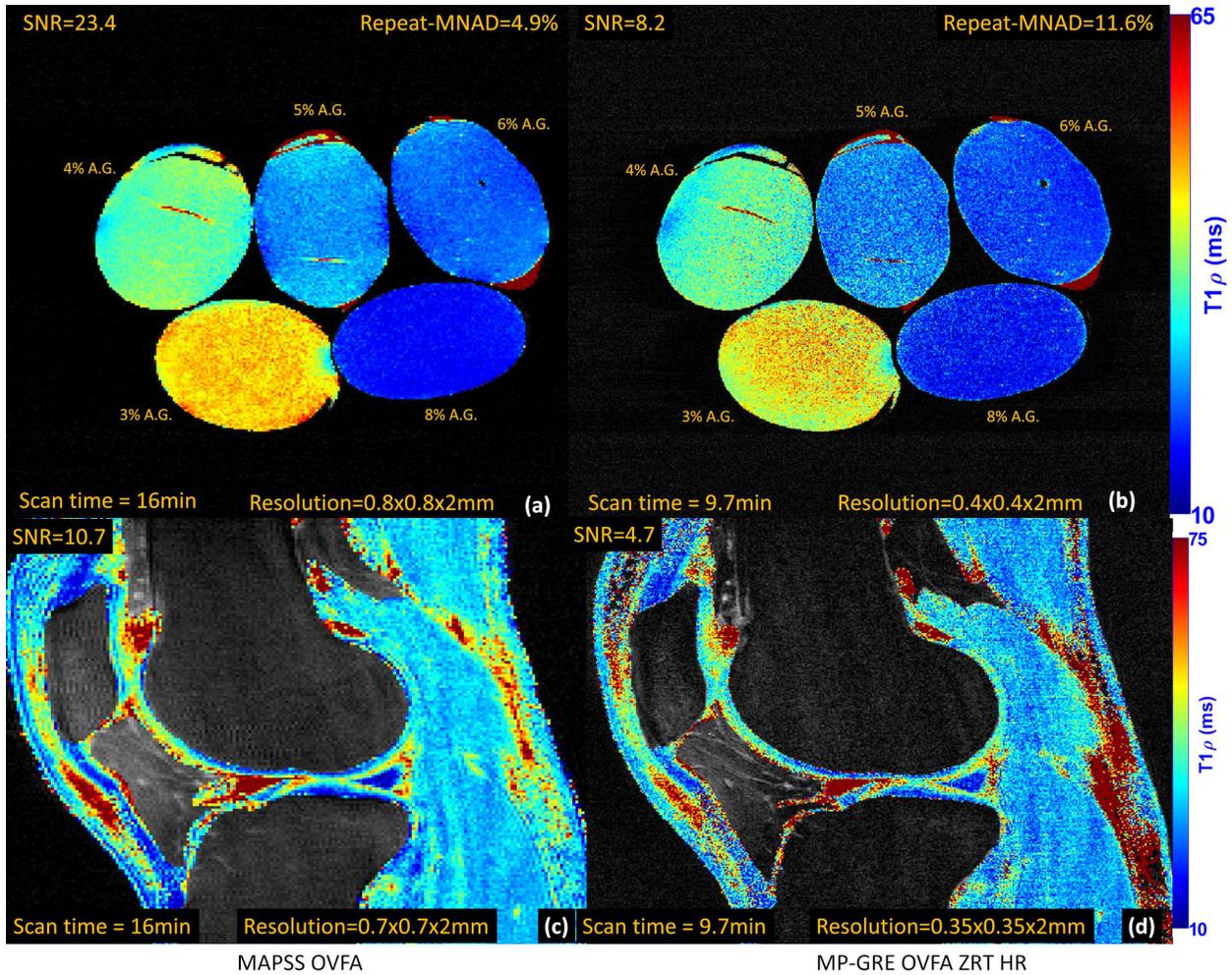

**Figure 8:** Illustration of the performance of the MP-GRE-OVFA-ZRT-HR sequence when compared with MAPSS-OVFA.

## 4. DISCUSSION AND CONCLUSION

The OVFA improved all three $T_{1\rho}$ mapping sequences. Perhaps the important target improvement was the SNR, which is relevant in the MAPSS due to the chosen voxel size since this sequence is already the most accurate of all three (see Figure 4). In MP-GRE and MP-GRE-WR, besides SNR, the improvement in accuracy was also very important, correcting the over-estimation of large $T_{1\rho}$ values in MP-GRE-WR and under-estimation in MP-GRE (see figures 5 and 6).

By changing the weighting parameters and $\boldsymbol{m}_{ref}$ in the cost function, the user can choose which aspect of a particular sequence he/she wants to improve (see Table 1). We illustrate this point on MP-GRE-OVFA ZRT sequences (see Figure 7). While there is substantial room for improvement in all aspects of the sequence, in the end, maximizing one aspect comes at the cost of reducing others. For example, increasing SNR may reduce the maximum flatness that can be achieved (see bold-letter results in Table 2).

Note that the methods presented in (26,28) are also optimized approaches. The proposed approach can obtain nearly the same results as them if we use only one set of relaxation parameters (K=1), and the weighting parameters are adjusted to a maximum flatness of the SE. In this sense, this proposed approach increases the flexibility of the optimization target, and also expands the OVFA to regular MP-GRE (without Mz reset).

Another interesting aspect is making the sequence faster, by using a combination of parameters that reduce the scan time. These changes usually cause a reduction in SNR and accuracy, but the OVFA can minimize this loss, achieving similar performance as a slower sequence. If time is still of concern, one can still include undersampling, particularly data-driven learned undersampling (18–22) and deep learning reconstructions (49–53). Assuming fully-sampled data is acquired, the reduction of scan time happens here due to acquiring more data per unit of time.

With OVFA, one can use previously unthinkable parameters in these sequences. One example is the MP-GRE ZRT (zero $T_1$ recovery time). The MP-GRE-OVFA-ZRT is very efficient because it does not have any idle time, as the others. Fully sampled data for $T_{1\rho}$ mapping can be acquired in just 3.5 min with the tested configuration (compared to 16 min with MAPSS) with good SNR. As an illustration of the possibilities of OVFA with this sequence, we were able to produce $T_{1\rho}$ maps with a voxel size 4× smaller with MP-GRE OVFA ZRT HR, obtaining fully-sampled high-resolution $T_{1\rho}$ maps in less than 10 min.

There are several aspects to be investigated and improved in the proposed optimization approach. Here the weighting parameters were chosen manually, by a systematic search that reduces MNAD in the simulated case. However, the automated procedure can perhaps obtain better choices, such as (54–56). The matrices used here were relatively simple. Other choices could potentially obtain better results.

Finally, we believe that this optimization approach can be extended to other kinds of sequences and other kinds of contrast, like $T_1$ and $T_2$ mapping (57,58), and different magnetic field strengths (59). The OVFA can make magnetization-prepared sequences with VFA more efficient, opening the possibility to achieve more accuracy, precision, speed, or improved spatial resolution in quantitative MRI.

# ACKNOWLEDGMENTS

This study was supported by NIH grants, R21-AR075259-01A1, R01-AR068966, R01-AR076328-01A1, R01-AR076985-01A1, and R01-AR078308-01A1 and was performed under the rubric of the Center of Advanced Imaging Innovation and Research (CAI2R), an NIBIB Biomedical Technology Resource Center (NIH P41-EB017183).